# Optically tunable microresonator using an azobenzene monolayer

Andre Kovach,[1*] Jinghan He,[2*] Patrick J.G. Saris,[2] Dongyu Chen,[3] and Andrea M. Armani[1,2,3a]

[1]Mork Family Department of Chemical Engineering and Materials Science, University of Southern California, Los Angeles, California 90089, USA

[2]Department of Chemistry, University of Southern California, Los Angeles, California 90089, USA

[3] Ming Hsieh Department of Electrical Engineering-Electrophysics, University of Southern California, Los Angeles, California 90089, USA

Photoswitchable organic molecules can undergo reversible structural changes with an external light stimulus. These optically controlled molecules have been used in the development of "smart" polymers, optical writing of grating films, and even controllable in-vivo drug release. Being the simplest class of photoswitches in terms of structure, azobenzenes have become the most ubiquitous, well-characterized, and implemented organic molecular switch. Given their predictable response, they are ideally suited to create an all-optically controlled switch. However, fabricating a monolithic optical device comprised solely from azobenzene while maintaining the photoswitching functionality is challenging. In this work, we combine integrated photonics with optically switchable organic molecules to create an optically controlled integrated device. A silica toroidal resonant cavity is functionalized with a monolayer of an azobenzene derivative. After functionalization, the loaded cavity Q is above $10^5$. When 450 nm light is coupled into cavity resonance, the azobenzene isomerizes from trans-isomer to cis-isomer, inducing a refractive index change. Because the resonant wavelength of the cavity is governed by the index, the resonant wavelength changes in parallel. At the probe wavelength of 1300 nm, the wavelength shift is determined by the duration and intensity of the 450 nm light and the density of azobenzene functional groups on the device surface, providing multiple control mechanisms. Using this photoswitchable device, resonance frequency tuning as far as sixty percent of the cavity's free spectral range in the near-IR is demonstrated. The kinetics of the tuning are in agreement with spectroscopic and ellipsometry measurements coupled with finite element method calculations.

* Both authors contributed equally.
a) Corresponding author: armani@usc.edu

Optical resonant cavities are a fundamental element of on-chip integrated optical circuits, serving as amplifiers, filters, and buffers.[1-7] These devices have two key features that differentiate them from other components: the ability to isolate and to store pre-defined or resonant wavelengths ($\lambda_o$). In many cases, it is desirable to change the $\lambda_o$, for example, tuning an add-drop filter or encoding an optical signal. Because the $\lambda_o$ is governed, in part, by the device refractive index, a common strategy is to leverage the electro-optic effect.[8-10] However, many optical cavities are fabricated from materials like silica with low to negligible electro-optic coefficients. Additionally, while the electro-optic effect can achieve fast tuning over small wavelength ranges, performing large shifts, comparable to the free spectral range of the cavity is challenging. Recent efforts have explored using the thermo-optic effect or the photo-acoustic effect to accomplish larger range tuning.[11-15] However, these control mechanisms are susceptible to cross-talk with adjacent optical components, thus, decreasing the density of the optical circuit.

An alternative strategy can be found in photoswitchable or light-triggerable organic materials. This emerging class of materials has rapidly gained prominence in a range of fields including photonics[16-19], plasmonics[20,-22], electronics[23-25], and even drug delivery[26-28]. This broad impact is possible because of the diversity of mechanisms available for light-triggerable materials such as photoisomerization, photopolymerization, photo-induced ring opening, and even photo-oxidation. Importantly, many of these changes are reversible. Traditional photoswitchable small molecules include azobenzenes[29], spiropyrans[30], diarylethenes[31], and a whole host of their derivatives.

Being the simplest class of photoswitches in terms of structure, azobenzenes have become the most ubiquitous, well-characterized, and implemented organic molecular switch. Consisting of two benzene rings joined by a nitrogen-nitrogen double bond, azobenzenes undergo a light-mediated isomerization that is fully recoverable by heat. Azobenzenes will typically experience a trans to cis isomerization when exposed to UV or blue light and will revert to the thermodynamically preferred trans isomer through application of heat or isolation in a dark environment (Fig. 1 a, b). The specific triggering wavelength or absorption band of these transitions can be tuned by modifying the functional groups on either of the benzene rings, and the refractive index of the azobenzene is dependent on the isomer.

This control is notable because the fundamental mechanism for resonant wavelength tuning is governed by a combination of diameter change and refractive index change [32]:

$$\Delta\lambda = \lambda\left(\frac{\Delta n}{n} + \frac{\Delta R}{R}\right) \quad (1)$$

where Δλ is the change in resonant wavelength, λ is the resonant wavelength, Δn is the change in refractive index, ΔR is the change in cavity radius, n is the effective refractive index and R is the cavity radius. By analyzing the equation, it becomes evident that it is the relative, not absolute, change in Δn or ΔR that governs Δλ. Therefore, the larger the device size, the more difficult it is for ΔR to play a dominant role in inducing a wavelength change. In contrast, Δn is diameter independent. Previous work demonstrated photoswitching through geometric modulation (ΔR). However, this approach required extremely thick layers of azobenzene which degraded the Q factor. [33] Other interesting work has leveraged the switching ability of bacteriorhodopsin layers integrated onto silica microspheres to produce switches and even realize photonic circuitry [34,35,36].

In the present work, arrays of microtoroidal cavities are fabricated from $SiO_2$, and the surface is functionalized with a monolayer of azobenzene (Fig. 1c, d). To study the optically tunable resonator, photoswitching is initiated using either a 410 nm or a 450 nm laser source, while the 1300 nm resonant wavelength is monitored. The tuning response as well as the maximum shift possible are characterized and are compared with finite element method simulations which model the refractive index change due to the isomerization occurring on the surface.

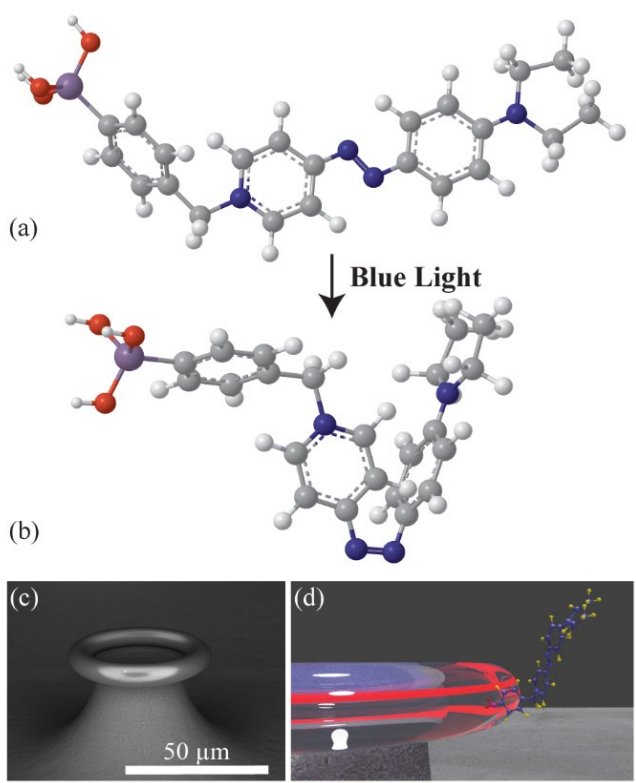

Fig. 1(a) Rendering of the trans state of the aminoazobenzene derivative free molecule and (b) the cis state of the molecule. Blue light initiates the trans to cis conformation change. (c) SEM image of coated microtoroid. (d) Rendering of coated microtoroid, with the molecule enlarged for clarity.

High-Q $SiO_2$ toroidal microcavities are fabricated on intrinsic Si wafers using photolithography, etching, and laser reflow using previously detailed procedures.[37] For the $SiO_2$ devices, the initial 2.0 μm layer of $SiO_2$ is grown using a wet thermal oxidation process. First, the wafers are patterned using Shipley 1813 photoresist, and then etched using Transene buffered oxide etchant and undercut using xenon difluoride ($XeF_2$). Finally, they are reflowed using a Synrad $CO_2$ laser. The toroidal cavities have major diameters from 40–50 μm and minor diameters from 6–7 μm, depending on slight variations in the $XeF_2$ undercut and laser reflow conditions.

An overview of the process used to form the photo-responsive monolayers is shown in Fig. 2. The initial silica microcavities are first treated by $O_2$ plasma to generate a dense layer of hydroxyl groups on the silica surface. Then, a [4-(chloromethyl)phenyl]trichlorosilane (CPS) coupling agent (Sigma, 97%) is deposited on the surface of the plasma-treated silica microtoroids using chemical vapor deposition at room temperature for 8 min, yielding a CPS-grafted microcavity.[38] The 4-(diethylamino)azobenzene (Aazo, Sigma) solution in chloroform is first mixed in a 1:1 (v/v) with a trichloromethyl silane (TCMS, Sigma), which acts as a spacer molecule to give the Aazo molecules more space on the surface to isomerize without steric hinderance, then drop-casted onto the CPS-grafted microcavities to form a uniform, oriented layer of Aazo. The 1:1 ratio was found to be optimal in terms of achieving the maximum refractive index change and maintaining high quality factors along with device sensitivity. Faster switching can be achieved by lowering the overall steric hinderance to the Aazo molecules on the surface by increasing the amount of spacer molecules. This, however, will come at a cost of tuning range and sensitivity to input pump powers. The utility of spacer molecules to optimize isomerization and switching times has been studied previously [39,40].Then, the coated devices are heated to 110 °C under vacuum for 20 min. The devices are cooled to room temperature, rinsed thoroughly with chloroform and acetone, and dried under vacuum at 110 °C for 5 min, yielding a grafted ~2 nm Aazo layer on the surface.[41] Control samples consisting of Aazo-functionalized $SiO_2$ on Si wafer are prepared using the same procedure for spectroscopic analysis.

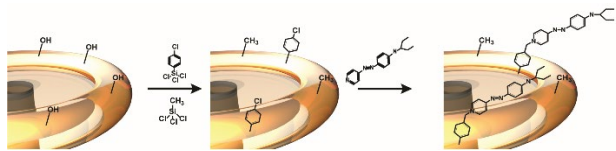

Fig. 2 Rendering of surface functionalization scheme showing attachment of the linker molecule (CPS) with subsequent Aazo attachment.

Several characterization measurements are performed on the Aazo compound before and after surface attachment to verify the photoswitchable behavior and the optical properties. The optical absorbance of a 39.5 µM Aazo solution in tetrahydrofuran is measured from 300 nm-1400 nm before and after exposure to 450 nm laser light (~1.6 W/cm$^2$). As seen in the spectra (Fig. 3(a)), the absorption maximum occurs at 450 nm and the absorption value decreases after exposure due to the π-π* transition of the molecule. [42] Additionally, there is minimal absorption in the near-IR in either state.

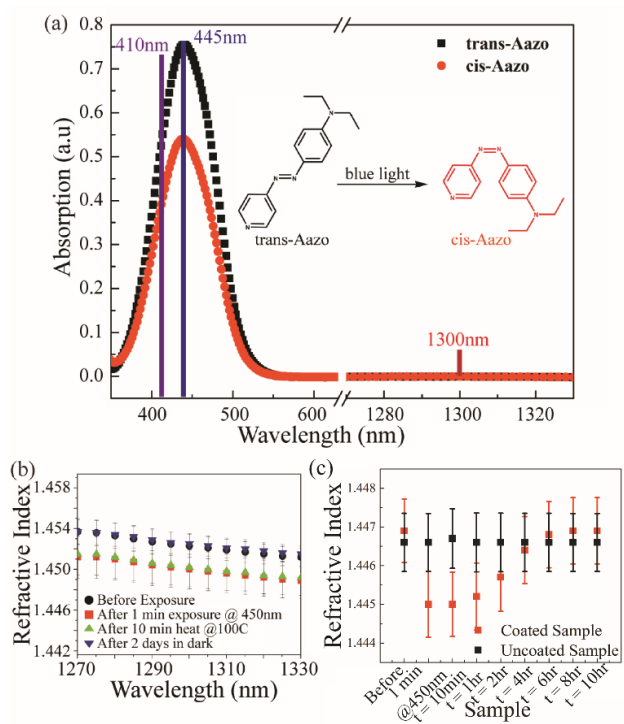

Fig. 3 (a) UV-Vis absorption spectra of both trans and cis isomer states. The wavelengths used to initiate photo-isomerization in the present work are indicated by vertical lines. (b) Spectroscopic ellipsometry results over the spectral range of the 1300 nm probe laser. (c) Measured refractive index for each step: before exposure, after exposure, and after heating for different time periods.

The covalent attachment and optical response of the Aazo layer is characterized using two methods. The surface chemistry is first confirmed by X-Ray photoelectron spectroscopy (XPS). The appearance of a chlorine peak in the CPS-functionalized wafer confirms the silanization reaction. The appearance of a nitrogen-associated peak in the Aazo functionalized sample confirms the Aazo attachment. The optical response of the bound Aazo layer is analyzed using spectroscopic ellipsometry from 1000–1700nm (V-VASE, J.A. Woollam Co.). The refractive index is determined using a five parameter Sellmeier equation and the fits agree well with literature. The results, zoomed in around the probe wavelength of 1300 nm, are shown in Figure 3(b). The scans show a 0.0023 decrease in the refractive index of the control wafers after exposure to 450 nm light, indicating a trans to cis isomerization.

Furthermore, the time dependent kinetics were also measured using ellipsometry on a separate coated and uncoated silica wafer (Figure 3(c)). The refractive index at 1300 nm was measured before and after a one-minute exposure to 450 nm laser light and at distinct time points with heat being applied via a hotplate. As can be seen, adding heat accelerates the recovery process, reducing the time to full recovery to about 6-8 hours from over 24 hrs without heat, it is still a considerably slow process. As expected, an uncoated silica control wafer showed little to no variation in refractive index when undergoing the same thermal treatments. Other methods to accelerate the reverse isomerization process were also explored, but either were not suitable for combination with integrated devices or contained intrinsic confounding complicating data analysis.

To calculate a theoretical resonant wavelength shift, both the ΔR and Δn terms must be considered. ΔR is simply the change in the Aazo length (0.3-0.5nm). Thus, the ΔR/R term is between 1E-5 to 3E-5. To calculate the index change, density functional theory (DFT) modeling of the molecular layer must be combined with finite element method modeling results of the optical mode. This approach allows for the effective refractive index ($n_{eff}$) term in Eq (1) to be calculated according to:

$$n_{eff} = \alpha n_{Cavity} + \beta n_{Layer} + \gamma n_{Air} \qquad (2)$$

where α, β, and γ represent the portions of the optical field inside the cavity, inside the layer of Aazo, and outside the cavity, respectively. [43] These values can be obtained using finite element method (FEM) simulations. The refractive index of the layer in the cis and trans states is approximated using density functional theory (DFT), and the value for silica is taken from the COMSOL library.

To accurately perform this modeling, the precise device geometries used in the experimental component of this work are used as well as the specific wavelengths. Additionally, cavities with varying major and minor radii are modeled in order to determine any significant changes in mode volumes due to the isomerization, as seen in Figs 4(c)-(f). The total mode volume values largely agree with previously published reports. [44] As expected, there is not a large change in total mode volume due to the change in the thin molecular layer. However, there is an appreciable change in the mode volume within the thin, higher index layer. In the trans state, more of the optical mode resides in the thin layer due to the larger refractive index, Figs 4(d) and (f). However, the actual change is small compared to that of the entire cavity system, ensuring that the majority of the optical mode remains relatively unperturbed, allowing the device to maintain a high Q. Based on these values, it is anticipated that the $\Delta n_{eff}/n$ term will vary between $4.1 \times 10^{-4}$ and $5.6 \times 10^{-4}$.

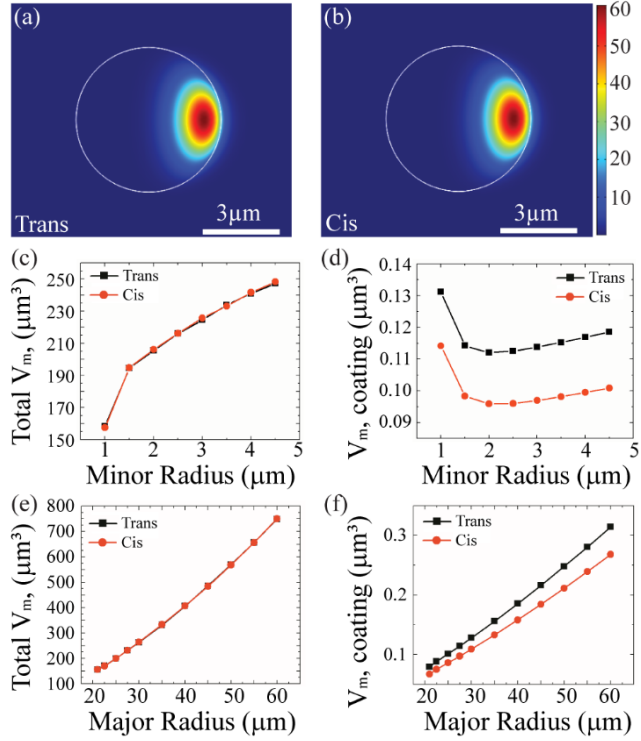

Fig. 4 (a), (b) COMSOL plots of optical mode profiles of both trans and cis states for a device with major and minor diameters of 50 µm and 7 µm. (c) Simulated total optical mode volume and (d) mode volume in the thin Aazo film as a function of minor radii for both cis and trans states. The major radius was fixed at 25 µm. (e) Simulated total optical mode volume and (f) mode volume in the thin film as a function of major radii for both cis and trans states. The minor radius was fixed at 3.5 µm. For a given geometry, the trans state has a larger interaction with the optical field than the cis state.

To characterize the tunability and determine the quality factor, light from a pair of lasers is coupled into the cavity using a single tapered optical fiber waveguide (Figure 5(a)). The configuration is similar to a classic pump-probe measurement set-up; however, the same tapered fiber (Newport F-SMF-28) couples both lasers into the cavity. The fiber is tapered down to roughly 600 nm in order to allow enough of the shorter wavelength (410 nm or 450 nm) light couple into the device when in contact, while also minimizing power loss of the longer wavelength (1300 nm) light. It is important to note that two different coupling methods are used simultaneously. Scattering is used to couple in the shorter wavelength light while evanescent coupling is used for the longer wavelength. [45] As a result, the coupling efficiencies are different. Approximately, 15% of the 410 nm or 450 nm light is coupled, and about 30-40% of the 1300 nm light is likely evanescently coupled into the toroid.

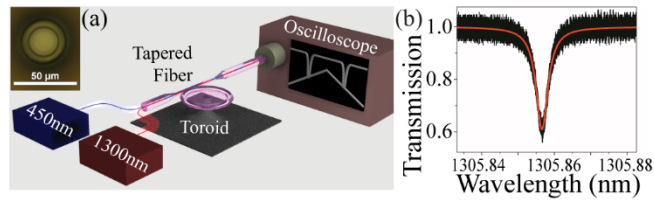

Fig. 5(a) Rendering of the characterization setup. (b) Representative transmission spectrum of a coated microtoroid on resonance at 1300nm (Q ~ 4.4x10$^6$).

Two different pump lasers are used to photoswitch the Aazo: a 410 nm narrow linewidth tunable laser (Newport, Velocity series) and a 450 nm diode laser (Thor Labs). This pair of lasers allowed several effects to be investigated. While both sources fall within the absorption range of the Aazo group, the 450 nm is at the maximum absorption of the Aazo. Additionally, the 410 nm is a narrow-linewidth source, allowing for investigation of resonant-enhancement of the index change. Specifically, in one measurement, both the 410 nm and the 1300 nm lasers were tuned to be simultaneously on-resonance. In the second measurement, only the 1300 nm laser was on-resonance. In the case of the 450 nm laser, the laser linewidth is very large (roughly 1000 times) compared to the linewidth of the cavity being tested, and the resonant cavity free-spectral range (FSR) is also small at 450 nm (~1.6 nm). As such, the laser is statistically likely to match a resonant wavelength and continue to do so even as the refractive index of the cavity changes. Therefore, measurements acquired using the 450 nm laser represent the maximum shift expected.

The probe laser is a 1300 nm narrow linewidth tunable laser (Newport, Velocity series). The 410 nm or 450 nm laser initiates the trans to cis isomerization on the surface while the 1300 nm tunable laser sweeps across and tracks the change in the cavity resonance. The alignment of the waveguide with the resonator is optimized using 3-axis nanopositioning stages and is monitored using top and side view cameras. The output signal is detected using a photodetector that is connected to a LabView-controlled high-speed digitizer/oscilloscope. [46] By scanning across a series of wavelengths, the resonant wavelength is identified, and the spectrum is recorded. By fitting this spectrum to a Lorentzian, the linewidth ($\delta\lambda$) is determined, and the loaded cavity Q at either 410 nm or 1300 nm is calculated. Typical loaded quality factors of the functionalized devices at 410 nm are in the mid-high 10$^5$. To determine the intrinsic cavity Q at 1300 nm, the amount of coupled power is varied, and a coupled cavity model is used. [47] Typical intrinsic Q factors at 1300 nm are in the mid-10$^6$ range (Fig. 5(b)). To minimize non-linear effects which may distort the spectrum and any subsequent analysis, the input power from the 1300 nm laser is kept below approximately 10 µW, and the coupling is kept relatively constant. The power being coupled into the cavity from the 410 nm laser is ~200 µW and from the 450 nm laser is ~1.5 mW.

Two different approaches are used to monitor the resonant wavelength. The first method relies on the tunable laser's ability to continuously raster across a narrow wavelength range, allowing for real-time monitoring and tracking of the peak position and transmission. This method allows for a single resonance peak to be monitored with high temporal and spatial resolution [48]. However, the precision comes at the cost of working range, and large shifts are unable to be measured. The second approach solves this challenge by leveraging the broad tunability of the 1300 nm laser. By scanning over the entire wavelength range of the laser, the entire resonance spectrum can be recorded, and all resonant peaks of the cavity identified and tracked. However, while this method solves the challenge associated with working range, the temporal and spatial resolution are reduced. Therefore, the majority of results are acquired using the first method.

The results from all photo-switching measurements are in Figure 6. Three measurements are performed while continuously rastering the 1300 nm laser over a single resonance while a fourth data set is acquired by analyzing a series of broadband transmission spectra. There are several points worthy of note. First, without the Aazo functional group, the resonant wavelength does not shift within the noise of the measurement. This stability indicates that any red or blue shifts measured in Aazo-functionalized devices can be attributed to the presence of the Aazo, and no other competing optical effects, such as the thermo-optic effect.

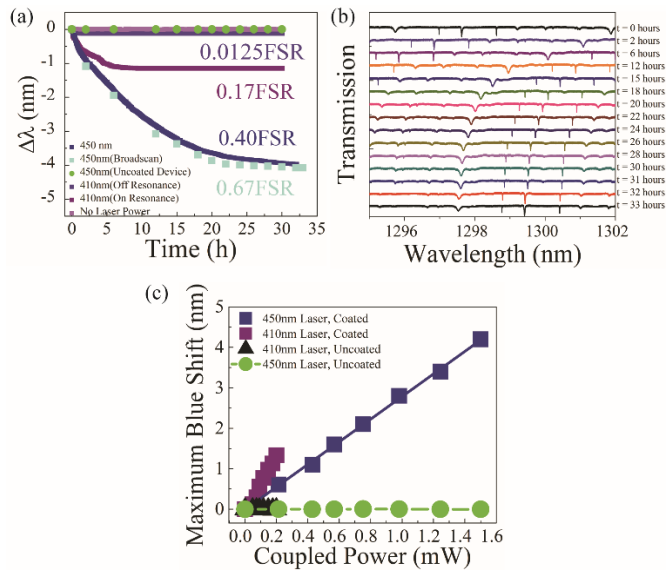

Fig. 6(a) Real-time tracking of the 1300 nm resonance peak tuning due to photo-isomerization upon exposure to either the 410 nm or 450 nm laser. (b) Broadband transmission spectra tracking the shifting resonance peak. (c) Sensitivity analysis using of coated device with varying input powers for both the 410 nm and 450 nm pumps

When the 410 nm laser was tuned off-resonance with 200 µW of input power, a small fraction of blue light was still scattered into the Aazo functionalized cavity, and a small probe resonant wavelength shift of 180 pm was observed (Fig. 6(a)). In contrast, if the 410 nm was tuned on-resonance, the 1300 nm resonance shift increased by nearly an order of magnitude from 180 pm to 1.3 nm. This increase is due to the resonant enhancement of the 410 nm light intensity. However, the 410 nm laser does not coincide with the absorption maximum of the Aazo molecule, limiting the full extent to which isomerization can take place. As a result, the maximum observable shift is about 1.3 nm, which corresponds to 0.17 FSR, even with Q's in the high $10^5$ range at 410 nm.

To overcome this limitation, a different testing configuration is used. The 410 nm narrow linewidth tunable laser is replaced with a 450 nm diode laser (~1 nm FWHM). This approach allows for continuous pumping of the resonance wavelength. However, using this fixed wavelength laser system, it is not possible to measure the cavity Q at 450 nm. With this change, the shift increases to 4 nm, which is roughly a half order of magnitude further improvement. This additional increase is likely due to the higher optical intensity of the 450 nm laser, the higher optical absorption of the azobenzene at 450 nm as compared to 410 nm, and the ability to maintain resonant excitation. This shift corresponds to about 40% of that specific cavity's FSR at the probe wavelength (~1300 nm).

To further verify these measurements, a series of broadband spectra are acquired, and the wavelength shift of a single peak is calculated. Characteristic results are shown in Fig. 6 (b). Over the course of about 33 hours the resonance peaks shift about 4.2 nm, which corresponds to about 0.67 FSR of that device, with about 1 nm of tuning within the first two hours. This represents a large contrast in the switching behavior. While the resonance peak itself is about 5 pm wide at half maximum, the total tuning range is about three orders of magnitude larger. Although switching contrast in our devices is large, it is still limited by the quality of the coupling at the pump wavelength, which is roughly 15% as we are unable to achieve critical coupling. This presents an area of improvement whereby contrast and switching speed could likely be increased by improvements in coupling through waveguide engineering. Broadly scanned resonance data, plotted alongside the other peak tracking results in Fig. 6 (a), largely agree and present very similar kinetics with each other. As mentioned, the inherent disadvantage of this method presents itself as poor spatial and temporal resolution as compared to continuously tracking the resonance peak. Nevertheless, this method provides an important validation on continuous peak tracking.

Using equation (1) and assuming that the radius change is negligible, the approximate refractive index change was calculated based on the resonant wavelength shift results and compared to the refractive index change measured via ellipsometry. The resonance shift yields a refractive index change of ~0.0028 which is roughly the refractive index change measured by ellipsometry of ~0.0023. Any discrepancy is most likely due to the much longer exposure during resonant cavity testing and the resonantly enhanced exposure to the 450 nm laser light as well as error in the spectroscopic ellipsometry measurement.

Analysis of the noise of our time dependent measurement was done to gauge the overall effect noise from the pump lasers contributes to the probe sensing. While coupling 410 nm or 450 nm light into the cavity does increase the noise of the 1300 nm peak shift measurement slightly, the noise level is always similar to or less than the linewidth of the cavity, with a maximum three-sigma noise level of around 3.4 pm. This value is far below the photoisomerization-induced resonance peak shift, which is three orders of magnitude larger. Additionally, the noise distribution is Gaussian, indicating that the cause of the noise is not due to any non-linear optical processes.

As shown in Fig. 6(a), the overall temporal response of the Aazo-coated device is, as expected, non-linear. This response mirrors the kinetics of the molecule in solution which are also temporally non-linear, albeit with a much faster response. The slower response of the Aazo when grafted onto a surface, in comparison to a liquid environment, is expected since the motion of the molecule is restricted. The grafting anchors half of the molecule's possible isomer transitions and steric hinderance can play a role due to closely packed neighboring molecules. In solution, isomerizations can usually take place in a few minutes, given favorable solvent conditions and low enough solute concentrations. However, while the tuning speed is slow, it compensates for this performance in the range of tunability and power sensitivity.

Depending on the input wavelength, the device response is governed by a balance of device Q and Aazo absorption. For example, at 410 nm, the tuning sensitivity is roughly 7.93 nm/mW of input power. This level of optical sensitivity is likely only achieved through the resonantly enhanced absorption of blue light within the device. By moving to 450 nm, where the optical absorption of the Aazo is higher, the sensitivity actually decreases to 2.79 nm/mW of input power. This decrease is most likely due to poor excitation of optical modes. The device shows a linear response at both wavelengths, indicating no competing non-linear effects. Moreover, the lasers are limited in their max power output and therefore limits our ability to observe any saturation of sensitivity. Although the cavity can be pumped by the 450 nm laser with an order of magnitude more power than the 410 nm laser, it possesses a much larger linewidth as stated previously. This will likely cause multiple resonances to be excited within the spectral bandwidth of the laser decreasing the efficiency of excitation.

Lastly, the surface chemistry demonstrated in this study creates a robust photoswitch consisting of covalently attached azo moieties. All data taken in Fig. 6 (b) was performed on the same device, repeatedly insulting the device with high intensity circulating optical fields. For reference, the circulating intensities in a device used in this work at 410 nm can reach roughly 9 MW/cm$^2$ given a quality factor of about 7.7x10$^5$. The linear response indicates that the Aazo groups did not degrade throughout the numerous cycles, despite these high optical fields. This stability is to be expected given that the thermal damage threshold is essentially dictated by the strength of the linking siloxy bond on the surface. Thermal degradation of a typical siloxy bond begins around 200°C [49,50]. Thermal degradation of silica, by comparison, begins around 1100°C, making our overall device system very robust.

In conclusion, we have demonstrated a method for broadband resonant cavity tuning via a covalently attached monolayer of a photoresponsive azobenzene derivative. The advantages of this method include robust attachment of a simple surface chemistry that does not change the operability or optical performance of the devices. This method also allows for broad tunability and fine control with fast tunability within the first minute of isomerization, yielding a maximum of 4.2 nm (0.67 FSR) of tuning. This wavelength range can be useful for applications where all-optical devices are desirable, such as optical add/drop filters, atomic clocks, and optomechanics. [51,52] In addition, the sensitivity of the surface chemistry to a range of input powers is characterized. As part of this analysis, the Aazo group is cycled multiple times using the natural relaxation intrinsic to the Aazo molecule. Future work could focus on improving reverse isomerization methods through the application of heat [53], improving the on-chip coupling methods to increase the switching contrast, and applying these findings to on-chip switchable microlasers and optically tunable on-chip frequency combs by combining with other nonlinear organic materials. [54,55] Overall, the outlook for integrating switchable organic molecules with WGM microresonators is promising.

The authors would like to thank Mark Veksler (University of Southern California) for renderings. This work is supported by the Army Research Office (ARO) (W911NF1810033) and Office of Naval Research (ONR) (N00014-17-2270).